\documentstyle[12pt]{article}

\begin{document}

\title{Relativistic Newton and Coulomb Laws}

\author{Yury M. Zinoviev\thanks{This work was supported in part by
the Russian Foundation for Basic Research (Grant No. 07 - 01 -
00144) and Scientific Schools 672.2006.1.}}

\date{}
\maketitle

Steklov Mathematical Institute, Gubkin St. 8, 119991, Moscow, Russia,

 e - mail: zinoviev@mi.ras.ru

\vskip 1cm

\noindent {\bf Abstract.} The relativistic equations for the
electromagnetic and gravitation interactions are similar: The only
Lagrangian equation is the equation with Lorentz force. The
potential satisfies the wave equation with the right - hand side
proportional to the velocity of another particle multiplied by the
delta - function concentrated at the position of another particle.
If the interaction propagates at the speed of light, then the wave
equation has the unique solution: the Li\'enard - Wiechert
potential. The Maxwell equations are completely defined by the
obtained relativistic Coulomb law. The Coulomb law and the Newton
gravity law differ from each other only in the choice of the
constants. If we choose in Coulomb law the electric charges equal to
the masses and choose the interaction constant of another sign, then
we get Newton gravity law. If we choose in the relativistic Coulomb
law the electric charges equal to the masses and choose the
interaction constant of another sign, then we get the relativistic
Newton gravity law.

\vskip 1cm

\section{Introduction}
\setcounter{equation}{0}

The celestial mechanics is based on the gravity law discovered by
Newton (1687). Cavendish (1773) proved by experiment that the force
of interaction between the electric charged bodies is inversely
proportional to the square of distance. This discovery was left
unpublished and later was repeated by de Coulomb (1785). The
electrodynamics equations were formulated by Maxwell (1873).  The
analysis of these equations led Lorentz \cite{1}, Poincar\'e
\cite{2} (this paper is the short version of the paper \cite{4}),
Einstein \cite{3} and Minkowski \cite{5} to the creation of the
theory of relativity. The Lorentz paper \cite{1} was based on the
covariance of Maxwell equations under the Lorentz transformations.
The Lorentz transformation and the Lorentz group were correctly
defined by Poincar\'e \cite{2}, \cite{4}. The Lorentz proof \cite{1}
of the covariance of Maxwell equations under the Lorentz group was
also corrected by Poincar\'e \cite{2}, \cite{4}. The idea of the
relativistic Newton gravity law was proposed by Poincar\'e \cite{4}:

"In the paper cited Lorentz \cite{1} found it necessary to
supplement his hypothesis in such a way that the relativity
postulate could be valid for other forces in addition to the
electromagnetic ones. According to his idea, because of the Lorentz
transformation (and therefore because of the translational
movement), all forces behave like electromagnetic (despite their
origin).

"It turned out to be necessary to consider this hypothesis more
attentively and to study the changes it makes in the gravity laws in
particular. First, it obviously enables us to suppose that the
gravity forces propagate not instantly but at the speed of light.
One could think that this is a sufficient for rejecting such a
hypothesis, because Laplace has shown that this cannot occur. But,
in fact, the effect of this propagation is largely balanced by some
other circumstance; hence, there is no any contradiction between the
law proposed and the astronomical observations.

"Is it possible to find a law satisfying the condition stated by
Lorentz and at the same time reducing to the Newton law in all the
cases where the velocities of the celestial bodies are small to
neglect their squares (and also the products of the accelerations
and the distance) compared with the square of the speed of light?"

This problem was formulated in other words in the paper \cite{2}.
The general relativity \cite{6} was another attempt to solve the
gravity problem. Einstein tried hard all his life to unify the
general relativity and the electrodynamics. It seems quite natural,
for the Coulomb law and the Newton gravity law differ from each
other only in the choice of the constants. In this paper we find the
relativistic Coulomb law by making use of the Poincar\'e
requirements and the additional requirements. The Maxwell equations
are completely defined by the obtained relativistic Coulomb law. If
we choose in the relativistic Coulomb law the electric charges equal
to the masses and choose the interaction constant of another sign,
then we get the relativistic Newton gravity law.

\section{Relativistic laws}
\setcounter{equation}{0}

We look for the relativistic Coulomb law in the following form
\begin{equation}
\label{1.17} mc\frac{dt}{ds} \frac{d}{dt} \left( \frac{dt}{ds}
\frac{dx^{\mu}}{dt} \right) + \frac{q}{c} \sum_{k\, =\, 0}^{N}
\sum_{\alpha_{1}, ..., \alpha_{k} \, =\, 0}^{3} \eta^{\mu \mu}
F_{\mu \alpha_{1} \cdots \alpha_{k}}(x)\frac{dt}{ds}
\frac{dx^{\alpha_{1}}}{dt} \cdots \frac{dt}{ds}
\frac{dx^{\alpha_{k}}}{dt} = 0,
\end{equation}
\begin{equation}
\label{1.18} \frac{dt}{ds} = c^{- 1}\left( 1 - c^{- 2}\Bigl|
\frac{d{\bf x}}{dt}\Bigr|^{2}\right)^{- 1/2}
\end{equation}
where $x^{0} = ct$, $\mu = 0,...,3$ and the diagonal $4\times 4$ -
matrix $\eta^{\mu \nu} = \eta_{\mu \nu}$, $\eta^{00} = - \eta^{11} =
- \eta^{22} = - \eta^{33} = 1$. The definition (\ref{1.18}) implies
the identities
\begin{eqnarray}
\label{1.19} \sum_{\alpha \, =\, 0}^{3} \eta_{\alpha \alpha} \left(
\frac{dt}{ds} \frac{dx^{\alpha}}{dt} \right)^{2} = 1, \nonumber \\
\sum_{\alpha \, =\, 0}^{3} \eta_{\alpha \alpha} \frac{dt}{ds}
\frac{dx^{\alpha}}{dt} \frac{dt}{ds} \frac{d}{dt} \left(
\frac{dt}{ds} \frac{dx^{\alpha}}{dt} \right) = 0.
\end{eqnarray}
The equation (\ref{1.17}) and the second identity (\ref{1.19}) imply
\begin{equation}
\label{1.20} \sum_{k\, =\, 0}^{N} \sum_{\alpha_{1}, ..., \alpha_{k +
1} \, =\, 0}^{3} F_{\alpha_{1} \cdots \alpha_{k + 1}}(x)
\frac{dt}{ds} \frac{dx^{\alpha_{1}}}{dt} \cdots \frac{dt}{ds}
\frac{dx^{\alpha_{k + 1}}}{dt} = 0.
\end{equation}
Let the functions $F_{\alpha_{1} \cdots \alpha_{k + 1}}(x)$ satisfy
the equation (\ref{1.20}). Then three equations (\ref{1.17}) for
$\mu = 1,2,3$ are independent
\begin{eqnarray}
\label{1.21} m\frac{d}{dt} \left( (1 - c^{- 2}|{\bf v}|^{2})^{-
1/2}v^{i}\right) - q\sum_{k\, =\, 0}^{N} c^{- k}(1 - c^{- 2}|{\bf
v}|^{2})^{- \frac{k - 1}{2}} \times \nonumber \\ \sum_{\alpha_{1},
..., \alpha_{k} \, =\, 0}^{3} F_{i\alpha_{1} \cdots
\alpha_{k}}(x)\frac{dx^{\alpha_{1}}}{dt} \cdots
\frac{dx^{\alpha_{k}}}{dt} = 0, \, \, v^{i} = \frac{dx^{i}}{dt},\,
\, i = 1,2,3.
\end{eqnarray}

The following lemma is proved in the paper \cite{7}.

\noindent {\bf Lemma 1}. {\it Let there exist a Lagrange function}
$L({\bf x},{\bf v},t)$ {\it such that for any world line},
$x^{\mu}(t)$, $x^{0}(t) = ct$, {\it and for any} $i = 1,2,3$ {\it
the relation}
\begin{eqnarray}
\label{1.22} \frac{d}{dt} \frac{\partial L}{\partial v^{i}} -
\frac{\partial L}{\partial x^{i}} = m\frac{d}{dt} \left( (1 - c^{-
2}|{\bf v}|^{2})^{- 1/2}v^{i}\right) - q\sum_{k\, =\, 0}^{N}
\nonumber \\ c^{- k}(1 - c^{- 2}|{\bf v}|^{2})^{- \frac{k - 1}{2}}
\sum_{\alpha_{1}, ..., \alpha_{k} \, =\, 0}^{3}F_{i\alpha_{1} \cdots
\alpha_{k}}(x)\frac{dx^{\alpha_{1}}}{dt} \cdots
\frac{dx^{\alpha_{k}}}{dt}
\end{eqnarray}
{\it holds. Then the Lagrange function has the form}
\begin{equation}
\label{1.23} L({\bf x},{\bf v},t) = - mc^{2}(1 - c^{- 2}|{\bf
v}|^{2})^{1/2} + \frac{q}{c} \sum_{i\, =\, 1}^{3} A_{i}({\bf
x},t)v^{i} + qA_{0}({\bf x},t)
\end{equation}
{\it and the coefficients in the equations} (\ref{1.21}) {\it are}
\begin{equation}
\label{1.24} F_{i\alpha_{1} \cdots \alpha_{k}}(x) = 0,\, \, k \neq
1,\, i = 1,2,3,\, \alpha_{1}, ...,\alpha_{k} = 0,...,3,
\end{equation}
\begin{eqnarray}
\label{1.25} F_{ij}(x) = \frac{\partial A_{j}({\bf x},t)}{\partial
x^{i}} - \frac{\partial A_{i}({\bf x},t)}{\partial x^{j}}, \nonumber
\\ F_{i0}(x) = \frac{\partial A_{0}({\bf x},t)}{\partial
x^{i}} - \frac{1}{c} \frac{\partial A_{i}({\bf x},t)}{\partial t},\,
i,j = 1,2,3.
\end{eqnarray}

We define
\begin{equation}
\label{1.26} F_{00} = 0,\, \, F_{0i} = - F_{i0},\, \, i = 1,2,3.
\end{equation}
Then the identity
\begin{equation}
\label{1.27} \sum_{\alpha, \beta \, =\, 0}^{3} F_{\alpha
\beta}(x)\frac{dt}{ds} \frac{dx^{\alpha}}{dt} \frac{dt}{ds}
\frac{dx^{\beta}}{dt} = 0
\end{equation}
similar to the identity (\ref{1.20}) holds. By making use of the
second identity (\ref{1.19}) and the identity (\ref{1.27}) we can
rewrite the equation (\ref{1.21}) with the coefficients
(\ref{1.24}), (\ref{1.25}) as the equation with Lorentz force
\begin{equation}
\label{1.28} mc\frac{dt}{ds} \frac{d}{dt} \left( \frac{dt}{ds}
\frac{dx^{\mu}}{dt} \right) = - \frac{q}{c} \eta^{\mu \mu} \sum_{\nu
\, =\, 0}^{3} F_{\mu \nu}(x)\frac{dt}{ds} \frac{dx^{\nu}}{dt}, \, \,
\mu = 0,...,3.
\end{equation}
Here we use the coefficients (\ref{1.25}), (\ref{1.26}).

The Coulomb law has the form
\begin{equation}
\label{1.29} m_{k}\frac{d^{2}x_{k}^{i}}{dt^{2}} = q_{k}
\frac{\partial}{\partial x_{k}^{i}} U({\bf x}_{k},{\bf x}_{j}),
\end{equation}
\begin{equation}
\label{1.30} \sum_{i\, =\, 0}^{3} \left( \frac{\partial}{\partial
x_{k}^{i}} \right)^{2} U({\bf x}_{k},{\bf x}_{j}) = 4\pi
q_{j}K\delta ({\bf x}_{k} - {\bf x}_{j}),\, \, i = 1,2,3,\, \, k,j =
 1,2,\, \, k \neq j,
\end{equation}
where $q_{1}, q_{2}$ are the electric charges and $K$ is the
constant. The Newton gravity law is the equations (\ref{1.29}),
(\ref{1.30}) with the constants $q_{k} = m_{k}$, $k = 1,2$, $K = -
\, G$ where the gravitation constant $G = (6.673 \pm 0.003)\cdot
10^{- 11}m^{3}kg^{- 1}s^{- 2}$.

We define the relativistic Coulomb law
\begin{equation}
\label{1.31} m_{k}c\frac{dt}{ds_{k}} \frac{d}{dt} \left(
\frac{dt}{ds_{k}} \frac{dx_{k}^{\mu}}{dt} \right) = -
\frac{q_{k}}{c} \eta^{\mu \mu} \sum_{\nu \, =\, 0}^{3} F_{\mu
\nu}(x_{k},x_{j})\frac{dt}{ds_{k}} \frac{dx_{k}^{\nu}}{dt},
\end{equation}
\begin{equation}
\label{1.32} F_{\mu \nu}(x_{k},x_{j}) = \frac{\partial
A_{\nu}(x_{k},x_{j})}{\partial x_{k}^{\mu}} - \frac{\partial
A_{\mu}(x_{k},x_{j})}{\partial x_{k}^{\nu}},
\end{equation}
\begin{equation}
\label{1.33} (\partial_{x_{k}}, \partial_{x_{k}})
A_{\mu}(x_{k},x_{j}) = \eta_{\mu \mu} 4\pi q_{j}K\left(
\frac{d}{dx_{k}^{0}} x_{j}^{\mu} \left( c^{- 1}x_{k}^{0} \right)
\right) \delta \left( {\bf x}_{k} - {\bf x}_{j} \left( c^{- 1}
x_{k}^{0} \right) \right),
\end{equation}
\begin{equation}
\label{1.34} (\partial_{x}, \partial_{x}) = \sum_{\nu \, =\, 0}^{3}
\eta^{\nu \nu} \left( \frac{\partial}{\partial x^{\nu}} \right)^{2},
\end{equation}
$$
\frac{dt}{ds_{k}} = c^{- 1}\left( 1 - c^{- 2}\Bigl| \frac{d{\bf
x}_{k}}{dt}\Bigr|^{2}\right)^{- 1/2},\, \, \mu = 0,...,3,\, \,
x_{k}^{0}(t) = ct,\, \, j,k = 1,2,\, \, j \neq k.
$$
The right - hand sides of the equations (\ref{1.30}) and
(\ref{1.33}) differ from each other in the velocity of another
particle.

The following lemma is proved in the paper \cite{7}.

\noindent {\bf Lemma 2}. {\it If the world line}, $x_{j}^{\mu}(t)$,
$x_{j}^{0} = ct$, {\it satisfies the condition}
\begin{equation}
\label{1.35} \Bigl| \frac{d{\bf x}_{j}(t)}{dt}\Bigr| < c,
\end{equation}
{\it then for an arbitrary matrix} $\Lambda_{\nu}^{\mu}$ {\it from
the Lorentz group}
\begin{eqnarray}
\label{1.36} \sum_{\nu \, =\, 0}^{3} \Lambda_{\nu}^{\mu} \left(
\frac{d}{dx_{k}^{0}} x_{j}^{\nu} \left( c^{- 1}x_{k}^{0} \right)
\right) \delta \left( {\bf x}_{k} - {\bf x}_{j} \left( c^{- 1}
x_{k}^{0} \right) \right) = \nonumber \\ c^{- 1}\left(
\frac{d}{dt_{1}} (\Lambda x_{j})^{\mu} (t(t_{1})) \right) \delta (
(\overrightarrow{\Lambda x_{k}}) - (\overrightarrow{\Lambda x_{j}})
(t(t_{1})))\Bigr|_{t_{1}\, =\, \, c^{- 1}(\Lambda x_{k})^{0}}
\end{eqnarray}
{\it where the function} $t(t_{1})$ {\it is defined by the equation}
\begin{equation}
\label{1.37} ct_{1} = (\Lambda x_{j})^{0}(t(t_{1})) = \sum_{\nu \,
=\, 0}^{3} \Lambda_{\nu}^{0} x_{j}^{\nu}(t(t_{1})).
\end{equation}

If the condition (\ref{1.35}) for $j = 1,2$ is valid, the relations
(\ref{1.36}), (\ref{1.37}) imply the Lorentz covariance of the
equations (\ref{1.31}) - (\ref{1.33}). The Lorentz covariance of the
Maxwell equations will be the consequence of the Lorentz covariance
of the equations (\ref{1.31}) - (\ref{1.33}) and the Poincar\'e
requirement that the interaction propagates at the speed of light.

If a distribution $e_{0}(x) \in S^{\prime}({\bf R}^{4})$ satisfies
the equation
\begin{equation}
\label{1.38} - \, (\partial_{x}, \partial_{x}) e_{0}(x) = \delta
(x),
\end{equation}
then a distribution
\begin{equation}
\label{1.39}  A_{\mu}(x_{k},x_{j}) = -\, \eta_{\mu \mu} 4\pi
q_{j}K\int d^{4}ye_{0}(x_{k} - y)\left( \frac{d}{dy^{0}} x_{j}^{\mu}
\left( c^{- 1}y^{0} \right) \right) \delta \left( {\bf y} - {\bf
x}_{j} \left( c^{- 1}y^{0} \right) \right)
\end{equation}
satisfies the equation (\ref{1.33}). Due to the Poincar\'e
requirement the interaction propagates at the speed of light. It
means that a support of a distribution $e_{0}(x)$ lies in the
boundary of the upper light cone.

\noindent {\bf Lemma 3}. {\it If a distribution} $e_{0}(x)$ {\it has
a support in the closed upper light cone and satisfies the equation}
(\ref{1.38}), {\it then}
\begin{equation}
\label{1.40} e_{0}(x) = -\, (2\pi)^{- 1} \theta (x^{0})\delta
((x^{0})^{2} - |{\bf x}|^{2})
\end{equation}
{\it where the step function}
\begin{equation}
\label{1.41} \theta (x) = \left\{ {1, \hskip 0,5cm x \geq 0,} \atop
{0, \hskip 0,5cm x < 0.} \right.
\end{equation}
{\bf Proof.} Let the equation (\ref{1.38}) have two solutions
$e^{(1)}(x)$, $e^{(2)}(x)$ whose supports lie in the closed upper
light cone. In view of the convolution commutativity these solutions
coincide
$$
e^{(2)}(x) =  - \, (\partial_{x}, \partial_{x}) \int d^{4}ye^{(1)}(x
- y)e^{(2)}(y) = - \, (\partial_{x}, \partial_{x}) \int
d^{4}ye^{(2)}(x - y)e^{(1)}(y) = e^{(1)}(x).
$$
Due to (\cite{8}, Section 30) the distribution (\ref{1.40}) is the
solution of the equation (\ref{1.38}). The lemma is proved.

The support of the distribution (\ref{1.40}) lies in the boundary of
the upper light cone. The Poincar\'e requirement is fulfilled. The
distribution (\ref{1.40}) is Lorentz invariant.

The substitution of the distribution (\ref{1.40}) into the equality
(\ref{1.39}) yields
\begin{eqnarray}
\label{1.42} A_{\mu}(x_{k}, x_{j}) = \eta_{\mu \mu} q_{j}K\left(
\frac{d}{dt^{\prime}} x_{j}^{\mu}(t^{\prime})\right) \left( c|{\bf
x}_{k} - {\bf x}_{j}(t^{\prime})| - \sum_{i\, =\, 1}^{3} (x_{k}^{i}
- x_{j}^{i}(t^{\prime})) \frac{d}{dt^{\prime}}
x_{j}^{i}(t^{\prime})\right)^{- 1}, \nonumber \\ x_{k}^{0} -
ct^{\prime} = |{\bf x}_{k} - {\bf x}_{j}(t^{\prime})|.
\end{eqnarray}
The potentials (\ref{1.42}) were introduced by Li\'enard (1898) and
Wiechert (1900) as the generalizations of Coulomb potential.

If the velocities of bodies are small enough to neglect their
squares compared with the square $c^{2}$ of the speed of light and
the time $t^{\prime}$ in the equality (\ref{1.42}) is approximately
equal to $c^{- 1}x_{k}^{0}$, then
\begin{eqnarray}
\label{1.43} A_{0}(x_{k},x_{j}) \approx q_{j}K|{\bf x}_{k} - {\bf
x}_{j}(c^{- 1}x_{k}^{0})|^{- 1}, \nonumber \\ A_{i}(x_{k},x_{j})
\approx 0, \, \, i = 1,2,3, \nonumber \\ \frac{dt}{ds_{k}} \approx
c^{- 1}.
\end{eqnarray}
The substitution of the expressions (\ref{1.43}) into the equations
(\ref{1.31}), (\ref{1.32}) yields the Coulomb law (\ref{1.29}),
(\ref{1.30}). All Poincar\'e requirements are fulfilled. If we
insert the constants $q_{k} = m_{k}$, $k = 1,2$, $K = - \, G$ into
the equations (\ref{1.31}), (\ref{1.32}), (\ref{1.42}), then we
obtain the relativistic Newton gravity law.

The interaction between $n$ charged bodies is given by the equations
\begin{equation}
\label{1.44} m_{k}c\frac{dt}{ds_{k}} \frac{d}{dt} \left(
\frac{dt}{ds_{k}} \frac{dx_{k}^{\mu}}{dt} \right) = -
\frac{q_{k}}{c} \eta^{\mu \mu} \sum_{\nu \, =\, 0}^{3}
\frac{dt}{ds_{k}} \frac{dx_{k}^{\nu}}{dt} \sum_{j\, =\, 1,...,n,\, j
\neq k} F_{\mu \nu}(x_{k},x_{j})
\end{equation}
where $\mu = 0,...,3$, $k = 1,...,n$ and the functions $F_{\mu
\nu}(x_{k},x_{j})$ are given by the relations (\ref{1.32}),
(\ref{1.42}).

\noindent {\bf Lemma 4}. {\it The potentials} (\ref{1.42}) {\it
satisfy the equation}
\begin{equation}
\label{1.45} \sum_{\mu \, =\, 0}^{3} \eta^{\mu \mu}
\frac{\partial}{\partial x_{k}^{\mu}} A_{\mu}(x_{k},x_{j}) = 0.
\end{equation}
{\bf Proof}. Let us define the vector
\begin{equation}
\label{1.46} j^{\mu}(x_{k},x_{j}) = 4\pi q_{j}K\left(
\frac{d}{dx_{k}^{0}} x_{j}^{\mu} \left( c^{- 1}x_{k}^{0} \right)
\right) \delta \left( {\bf x}_{k} - {\bf x}_{j} \left( c^{- 1}
x_{k}^{0} \right) \right),\, \mu = 0,...,3.
\end{equation}
The world line $x_{j}^{\mu}(t)$ satisfies the condition
$x_{j}^{0}(t) = ct$. Hence the definition (\ref{1.46}) implies
\begin{eqnarray}
\label{1.47} \frac{\partial}{\partial x_{k}^{0}} j^{0}(x_{k},x_{j})
= - 4\pi q_{j}K\sum_{i\, =\, 1}^{3} \left( \frac{d}{dx_{k}^{0}}
x_{j}^{i} \left( c^{- 1}x_{k}^{0} \right) \right)
\frac{\partial}{\partial x_{k}^{i}} \delta \left( {\bf x}_{k} - {\bf
x}_{j} \left( c^{- 1} x_{k}^{0} \right) \right), \nonumber \\
\frac{\partial}{\partial x_{k}^{i}} j^{i}(x_{k},x_{j}) = 4\pi
q_{j}K\left( \frac{d}{dx_{k}^{0}} x_{j}^{i} \left( c^{- 1}x_{k}^{0}
\right) \right) \frac{\partial}{\partial x_{k}^{i}} \delta \left(
{\bf x}_{k} - {\bf x}_{j} \left( c^{- 1} x_{k}^{0} \right)
\right),\, i = 1,2,3.
\end{eqnarray}
The relations (\ref{1.47}) imply the continuity equation
\begin{equation}
\label{1.48} \sum_{\mu \, =\, 0}^{3} \frac{\partial}{\partial
x_{k}^{\mu}} j^{\mu}(x_{k},x_{j}) = 0.
\end{equation}
The definitions (\ref{1.39}), (\ref{1.46}) and the equation
(\ref{1.48}) imply the equality (\ref{1.45}). The lemma is proved.

\noindent {\bf Lemma 5}. {\it If the functions} $F_{\mu
\nu}(x_{k},x_{j})$ {\it are defined by the relations} (\ref{1.32}),
(\ref{1.42}), {\it then}
\begin{equation}
\label{1.49} \sum_{\mu \, =\, 0}^{3} \eta^{\mu \mu}
\frac{\partial}{\partial x_{k}^{\mu}} \sum_{j\, =\, 1,...,n,\, j
\neq k} F_{\mu \nu}(x_{k},x_{j}) = \eta_{\nu \nu} \sum_{j\, =\,
1,...,n,\, j \neq k} j^{\nu}(x_{k},x_{j}).
\end{equation}
{\bf Proof}. The definition (\ref{1.32}) and the equation
(\ref{1.45}) imply
\begin{equation}
\label{1.50} \sum_{\mu \, =\, 0}^{3} \eta^{\mu \mu}
\frac{\partial}{\partial x_{k}^{\mu}} F_{\mu \nu}(x_{k},x_{j}) =
\sum_{\mu \, =\, 0}^{3} \eta^{\mu \mu} \left(
\frac{\partial}{\partial x_{k}^{\mu}} \right)^{2}
A_{\nu}(x_{k},x_{j}).
\end{equation}
The potential $A_{\nu}(x_{k},x_{j})$ is defined by the relations
(\ref{1.39}), (\ref{1.40}). The distribution (\ref{1.40}) satisfies
the equation (\ref{1.38}). Hence the equalities (\ref{1.34}),
(\ref{1.39}), (\ref{1.46}), (\ref{1.50}) imply
\begin{equation}
\label{1.51} \sum_{\mu \, =\, 0}^{3} \eta^{\mu \mu}
\frac{\partial}{\partial x_{k}^{\mu}} F_{\mu \nu}(x_{k},x_{j}) =
\eta_{\nu \nu} j^{\nu}(x_{k},x_{j}).
\end{equation}
By summing up the equalities (\ref{1.51}) we obtain the equality
(\ref{1.49}). The lemma is proved.

By making use of the definition (\ref{1.32}) it is easy to verify
for different $\mu, \nu, \lambda$
\begin{equation}
\label{1.52} \frac{\partial}{\partial x_{k}^{\lambda}} F_{\mu
\nu}(x_{k},x_{j}) + \frac{\partial}{\partial x_{k}^{\mu}} F_{\nu
\lambda}(x_{k},x_{j}) + \frac{\partial}{\partial x_{k}^{\nu}}
F_{\lambda \mu}(x_{k},x_{j}) = 0.
\end{equation}
The relations (\ref{1.49}), (\ref{1.52}) imply that the fields in
the relativistic Coulomb law (\ref{1.44}) satisfy the Maxwell
equations with the current given by the right - hand side of the
equality (\ref{1.49}). Therefore the electrodynamics is completely
defined by the relativistic Coulomb law (\ref{1.32}), (\ref{1.42}),
(\ref{1.44}). The substitution $q_{j} = m_{j}$, $K = - G$ in these
equations yields the relativistic Newton gravity law.

Let us consider the equations (\ref{1.31}), (\ref{1.32}),
(\ref{1.42}) for the case $q_{1} = m_{1} = 0$, $q_{2} = m_{2}$, $K =
- G$. Then the second body moves freely. These equations were solved
in the paper \cite{7}. The solution is similar to that of Kepler
problem. It is a principal result. There is no similar result in the
general relativity. It seems that the mass of Mercury is small and
the Sun moves freely. It is possible to calculate the advance of
Mercury's perihelion for a hundred years. According to \cite{7}, it
turns out to be $7".18$. According to \cite{6} (Chap. 40, Sec. 40.5,
Appendix 40.3), the whole observed advance is $5599".74 \pm 0".41$
for a hundred years. It is possible to calculate the advance of
Mercury's perihelion caused by non - inertial system connected with
the Earth using the Newton gravity law. According to \cite{6} (Chap.
40, Sec. 40.5, Appendix 40.3), it turns out to be $5025".645 \pm
0".50$ for a hundred years. It is also possible to calculate the
advance of Mercury's perihelion caused by the gravity of other
planets using the Newton gravity law. According to \cite{6} (Chap.
40, Sec. 40.5, Appendix 40.3), it turns out to be $531".54 \pm
0".68$ for a hundred years. There exists a contradiction between the
remaining advance of Mercury's perihelion of $42".56$ for a hundred
years and the calculated advance $7".18$. It seems that the
experimental testing of the relativistic Newton gravity law
(\ref{1.32}), (\ref{1.42}), (\ref{1.44}), $q_{j} = m_{j}$, $K = - G$
will be possible only when the entire advance of Mercury's
perihelion for a hundred years is calculated using these equations.
The quantum effects may be also important.

\end{document}